# A Universal Raman Spectroscopic Framework for Defect Quantification in Mono-to-Multilayer Graphenic Materials: The Graphene Atlas

Kazunori Fujisawa [1,2,3]*, Bruno R. Carvalho [4], Pedro Venezuela [5], Cheon-Soo Kang [1], Yoong Ahm Kim [6], Takuya Hayashi [1,7], Mauricio Terrones [2,3,8]*

[1] Research Initiative for Supra-Materials, Shinshu University, 4-17-1 Wakasato, Nagano, 380-8553, Japan.

[2] Department of Physics, The Pennsylvania State University, University Park, Pennsylvania 16802, United States.

[3] Center for 2-dimensional and Layered Materials (2DLM), The Pennsylvania State University, University Park, Pennsylvania 16802, United States.

[4] Departamento de Física, Universidade Federal do Rio Grande do Norte, Natal, Rio Grande do Norte, RN 59078-970, Brazil

[5] Institute of Physics, Universidade Federal Fluminense, Niteroi, RJ, 24210-346, Brazil

[6] Department of Polymer Engineering & Graduate School, School of Polymer Science and Engineering, Faculty of Engineering, Chonnam National University, 77 Yongbong-ro, Buk-gu, Gwangju 61186, Republic of Korea.

[7] Faculty of Engineering, Shinshu University, 4-17-1 Wakasato, Nagano, 380-8553, Japan.

[8] Department of Chemistry and Department of Materials Science and Engineering, The Pennsylvania State University, University Park, Pennsylvania 16802, United States.

Corresponding Author

E-mail address:

fujisawa@endomoribu.shinshu-u.ac.jp (K. Fujisawa), mut11@psu.edu (M. Terrones)

**Abstract**

Point defects, though atomically small, significantly influence the properties of 2D materials. A general method for characterizing point defect density ($n_D$) in graphenic materials with arbitrary layer number ($n_L$) is currently lacking. Here, we introduce the *Graphene Atlas*, a non-destructive Raman spectroscopy-based framework for defect quantification in diverse graphenic systems. We demonstrate that the relative fractions of the double-resonance D and 2D Raman bands, which arise from competing scattering processes, exhibit a universal relationship with $n_D$, independent of $n_L$. Plotting Raman data on a plane defined by defect-related and layer number-related parameters enables a direct and quantitative determination of $n_D$ and $n_L$. This *Graphene Atlas* provides a transformative tool for real-time defect quantification in scalable manufacturing of graphenic materials, bridging fundamental research and industrial applications. This framework establishes a new standard for defect characterization of graphenic systems, facilitating their optimization for advanced technological applications.

**Keywords:** Graphene, Raman spectroscopy, Point defects, layer-independent defect characterization, Graphene Atlas

Graphenic materials have shown intriguing physico-chemical properties that drastically change depending on their layer number ($n_L$). Monolayer graphene is renowned for its high mobility, transparency, surface-to-volume ratio, and universal optical absorbance[1]. Few-layered counterparts (2 to 5 of $n_L$) find applications in conductive inks for printing electronics,[2] while twisted few-layered graphenic materials near the "magical angle" exhibits superconductivity.[3,4] This versatility makes graphenic materials attractive for advanced electronics[5], catalysis[6,7], and sensing[8,9].

Unlike graphitic materials, where 1D line defects dominate, 0D point defects significantly influence the performance of graphenic systems. Accurate performance evaluation, therefore, requires precise knowledge of the defect density ($n_D$). While traditional imaging-based techniques, such as transmission electron microscopy (TEM)[10,11] and scanning tunneling microscopy (STM)[12,13], provide atomic resolution but are impractical for large-scale analysis, especially in multilayer systems where overlapping atomic planes obscure defects. Therefore, a method capable of robustly quantifying defect density $n_D$ across varying layer number $n_L$ is indeed required.

Raman spectroscopy, a non-destructive and non-contact technique, is widely used to evaluate the defectiveness in $sp^2$-hybridized carbon materials.[14] The Raman spectrum of these $sp^2$ carbon materials features the first-order graphitic G-band ($E_{2g}$) and the disorder-related D-band (TO(K)), which arises from a one-defect-one-phonon double-resonance (DR) Raman process. Existing methods use intensity ratios ($I_D/I_G$) or their integrated intensity (area) ratios ($A_D/A_G$), with experimental calibration, to estimate the lateral crystalline size ($L_a$) in graphitic materials[14,15] and to correlate with defectiveness—either defect density ($n_D(cm^{-2}) = 10^{14}/\pi L_D^2$) or inter-defect distance ($L_D(nm)$)—in monolayer graphene[16,17]. However, in multilayer graphenic systems, the Raman band shape, position, and width are strongly influenced by $n_L$ and stacking order, rendering $I_D/I_G$ an unreliable indicator of $n_D$ when $n_L$ varies[18]. This limitation has hindered defect quantification in multilayer systems and necessitates a unified approach. While previous studies have acknowledged this complexity and the competition of the D- and 2D-bands with increasing $n_D$[19–21], a universal framework for defect quantification across multilayer graphene has remained elusive[18].

Here, we present the *Graphene Atlas*, a Raman spectroscopy-based model that quantitatively determines $n_D$ (cm$^{-2}$) and $L_D$ (nm) independent of $n_L$. By analyzing over ten-thousand Raman spectra and plotting the data on an $n_D$–$n_L$ plane spanned by the defect-related $A_D/A_G$ and layer number-related $A_{2D}/A_G$ axes, we reveal a universal relationship enabling precise defect quantification in multilayer systems. Boron-doped mono-to-few-layered graphene, prepared through thermal diffusion and mechanical exfoliation[22], served as a platform for validating this approach. The introduced boron atoms within graphene work as point defects similar to other point defects ($sp^3$-defects and vacancies, etc.)[20].

Much like the *Kataura* plot revolutionized our understanding of single-walled carbon nanotubes by mapping their electronic transitions to specific structural parameters (chirality and diameter)[23], the *Graphene Atlas* provides a fundamental framework for correlating the Raman spectra of graphenic materials to their key structural properties: defect density $n_D$ and layer number $n_L$. This universal and layer number-independent framework overcomes the limitations of existing techniques and offers a new standard for defect characterization applicable in scalable manufacturing processes. This methodology enables real-time, non-destructive quality control and standardizes defect density and layer number quantification protocols, thus facilitating material certification and commercialization.

**Boron-doping of graphite**

Traditional methods for introducing point defects into graphenic materials, such as particle irradiation[24] and surface oxidation[25], often affect only the top few layers of the material[18,24]. This limits their effectiveness in creating uniformly defective materials. In this work, boron atoms were incorporated as point defects into bulk graphite, followed by mechanical exfoliation to obtain mono-to-few layered graphenic materials in high quality. Our boron-doping approach facilitates a more homogeneous distribution of defects throughout the resulting flakes..

Boron doping was achieved by placing graphite flakes and a small portion of boric acid ($H_3BO_3$) separately in a graphite crucible (**Fig. 1a**) and annealing them at 2400 ºC for 30 minutes under argon flow in a graphite furnace[22]. This high-temperature treatment facilitates the thermal diffusion of boron atoms into the bulk graphite.

The presence of boron atoms in the graphite flake was confirmed using time-of-flight secondary ion mass spectrometry (ToF-SIMS; **Fig. 1b**). This technique allows the exclusive detection of ionized boron atoms ($^{10}B^+$ and $^{11}B^+$) due to boron's lighter mass compared to carbon, eliminating interference from carbon-based ion species. The continuous detection of ionized boron atoms in the depth profile (**Fig. 1c**) confirms boron diffusion beyond the surface layers.

To investigate sub-surface diffusion of the boron atoms, Raman mapping was performed on freshly prepared cross-sections of pristine graphite (denoted as 'Gt') and boron-doped graphite (denoted as 'BGt'; **Fig. 1d**). In BGt, a relatively high $I_D/I_G$ ratio, indicative of high defect density ($n_D$), was observed near the top and bottom surfaces of the flakes. The $I_D/I_G$ ratio exhibited variation over a depth of approximately 100 μm from the surface, demonstrating that boron atoms diffused from the surface with a gradual change in concentration. While the subsequent mechanical exfoliation extracts nm-thick flakes of graphenic materials from BGt with μm-scale boron atom concentration change, the out-of-plane $n_D$ variation in the resulting thin flake should be negligibly small. In contrast, the non-uniform in-plane $I_D/I_G$ ratio observed in the Raman map indicates the presence of a μm-scale in-plane variation in $n_D$.

**Raman spectra of boron-doped graphene**

After thoroughly characterizing the BGt, we mechanically exfoliated flakes to isolate boron-doped graphene (BGr) thin flakes on top of the $SiO_2$/Si substrate. This process randomly extracts BGr from BGt, thus thin BGr with various $n_D$ and $n_L$ can be obtained by repeating the process. **Figures 2a** and **2b** show Raman spectra collected from boron-doped monolayer graphene (1L-BGr) and bilayer graphene (2L-BGr) with various $n_D$, respectively. Although similar Raman spectral change was observed, the empirical equation proposed by Cançado *et al.*, connecting $n_D$ and $I_D/I_G$ is only applicable to monolayers. Thus, a quantitative comparison of $n_D$ between 1L-BGr and 2L-BGr is not currently possible.

Close examination of the Raman spectra from 1L- and 2L-BGr reveals a significant difference in the 2D-band. Due to the overlapping of the Dirac cones in the bilayer system, multiple DR Raman processes are possible for the 2D-band, leading to its asymmetric shape. The 2D-band shape further changes when $n_L$ increases[26]. Similarly, the shape of the D-band, also originating from an iTO phonon-related DR process, changes with increasing the $n_L$[26]. **Figures 2a** and **2b** indicate that for low $n_D$ (corresponding to low $I_D/I_G$), the 2D-band remains with a similar intensity/area when compared to non-doped graphene, whereas at high $n_D$ (corresponding to high $I_D/I_G$), the 2D-band intensity/area decreases. This behavior has already been discussed theoretically[21] and experimentally[20]. While the $I_{2D}/I_G$ has been used to indicate $n_L$, our result shows that increasing $n_D$ affects the $n_L$ estimation. When a traveling electron/hole collides with defects in the graphene lattice before they are scattered by an iTO phonon that contributes to the 2D-band, defect-scattered electron/hole generates the D-band instead. While both 2D- and D-band processes are competing, the relative fraction of the D-band within the total iTO-related DR process (D+2D band) should dictate $n_D$ in the graphenic system.

To establish a universal guide for $n_D$ analysis in graphenic systems, we focused on the D/2D band area ($A_D$ and $A_{2D}$, respectively) rather than their intensities ($I_D$ and $I_{2D}$) to encompass all possible DR Raman signals contributing to these bands. **Figures 2c** and **2d** show that the $A_D/A_{2D}$ changes almost linearly with $I_D/I_G$ in both 1L and 2L systems, demonstrating that defect-related information can be extractable without using information about the G-band. In contrast, the G-band area ($A_G$) changed sensitively to $n_L$ (**Extended Fig. 1**). Because the G-band is a first-order Raman mode, its intensity ($I_G$) and area ($A_G$) should increase proportionally to $n_L$[27], and thus be used to determine $n_L$.

To separate the effect of $n_D$ and $n_L$, Raman data from 1L-BGr and 2L-BGr were plotted over a plane spanned by defect density ($n_D$)-related $A_D/A_G$ axis and layer number ($n_L$)-related $A_{2D}/A_G$ axis (**Fig. 2e**). This $n_D$–$n_L$ plane distinctly separates the data points from 1L- and 2L-BGr. As discussed, $A_{2D}$ decreases with increasing $n_D$, while $A_D$ increases in both 1L-BGr and 2L-BGr. The negative slope of the lines appeared in the $n_L$–$n_D$ plane (**Fig. 2e**) indicates $n_D$ variation at a constant $n_L$. A similar Raman data pattern, exhibiting a negative slope, was observed in other BGr flakes with constant $n_L$ and varying $n_D$ (red data points in **Fig. 4b** and **Extended Fig. 2**).

**Pattern analysis of Raman data on $n_D$–$n_L$ plane**

To disentangle the complex Raman features of BGr with varying $n_L$ and $n_D$, we further investigated the behavior of Raman data points on the $n_D$–$n_L$ plane. **Figure 3a** shows an optical image of a BGr flake and corresponding Raman images created using the areas of the main Raman bands: $A_G$, $A_D$, and $A_{2D}$. These Raman images reveal that $A_D$ and $A_{2D}$ exhibit gradual changes over the flake and are not strongly influenced by $n_L$, whereas $A_G$ changes abruptly with optical contrast, indicating local $n_L$ variations. Therefore, scaling $A_D$ and $A_{2D}$ by $A_G$ effectively separates Raman features based on local $n_L$ (**Fig. 3b**). The color of the Raman data points in **Fig. 3b** corresponds to the G-band intensity ($I_G$), as shown in the histogram, to visually distinguish local $n_L$. This $n_D$–$n_L$ plane then describes the distribution of $n_D$ and $n_L$ within the Raman image. Similar patterns, with broadly dispersed data points, were observed in other BGr flakes with non-constant $n_L$ and $n_D$ (gray-colored data points in **Fig. 4b**, **Extended Fig. 3**, and **Fig. S1-S3**). In contrast, undoped graphene flakes show Raman data points clustered around $A_{2D}/A_G$ axis (**Extended Fig. 4**), and BGr flakes with constant $n_L$ and constant $n_D$ show a single spot on the $n_D$–$n_L$ plane (blue-colored data points in **Fig. 4b**, **Extended Fig. 5**, and **Fig. S4-S7**).

Using **Fig. 3b** as a model, arbitrary $n_L$ can be estimated using $A_D/A_G$ and $A_{2D}/A_G$ correlations. By applying a linear regression technique, the equation (only applicable to 532 nm laser excitation wavelength) below is obtained,

$$\frac{A_{2D}}{A_G} + 0.87 \frac{A_D}{A_G} = \frac{4.74}{n_L} + 0.81 \quad (1)$$

To further investigate the relationship between Raman features and layer number $n_L$, we analyzed another BGr flake (**Fig. 3c**). Atomic force microscopy (AFM) image revealed regions with 1-4 layers and a thick, bulk-like region (**Fig. 3d**). Single-point Raman spectra from these regions (**Fig. 3e**) showed changes in the D-band shape with increasing $n_L$, and the relative intensity of the D-band with respect to the G-band also decreased, which is a typical tendency.

A key observation was made when plotting data from a BGr flake with non-constant $n_L$ and "unknown $n_D$" on the $n_D$–$n_L$ plane: the data points formed a line with positive slope (gray line in **Fig. 3f**, yellow-colored data in **Fig. 4b**, **Extended Fig. 6**, and **Fig. S8-S19**). To understand the physical origin of this line, we performed DR Raman cross-section calculations for monolayer, bilayer, and trilayer graphene, as described in Ref. 21. These calculations, using a tight-binding model with a varying first-neighbor hopping parameter ($\delta t$) to represent defects, showed that the D-band intensity is proportional to $\alpha_{hopp} = n_D(\delta t)^2$. The calculated $A_D$ and $A_{2D}$ values for various $n_D$ rendered by $\alpha_{hopp}$ are shown in **Fig. 4a**. While the G-band could not be calculated with this method, we focused on the relative fraction of $A_D$ and $A_{2D}$. Notably, the crossing point of the $A_D$ and $A_{2D}$ relative fractions remained ~$10^{14}$ regardless of the $n_L$. This indicates that the $A_D$ and $A_{2D}$ ratio exhibits a universal dependence on $n_D$, independent of $n_L$. Therefore, the observed line in **Fig. 3f** represents an equi-$n_D$ line— a line of constant defect density.

Based on this finding, we propose a guide for analyzing $n_D$ and $n_L$ from Raman data, which we term the *Graphene Atlas* (**Fig. 4c**). This atlas provides a universal relationship between $n_D$ and $n_L$, independent of layer number, enabling straightforward extraction of defect information and simplifying the analysis of complex graphenic multilayer systems. Experimental calibration of the equi-$n_D$ line will be discussed in the next section.

**Experimental calibration of the equi-$n_D$ line.**

Different BGr single flakes exhibited equi-$n_D$ lines with varying slopes on the $n_D$–$n_L$ plane (**Fig. 5a**). Detailed Raman data for these flakes are shown in **Figs. S8-12**, **S15**, and **S17-18**. The data points in **Fig. 5a** are colored according to their G-band intensity, indicating varying $n_L$. Because all equi-$n_D$ lines intersect the monolayer region, we can determine the defect density ($n_D$, in cm$^{-2}$) and the average distance between defects ($L_D$, in nm) corresponding to each slope using the empirical equation proposed by Cançado, *et al*[17]. Although the equation has been established based on graphene with topological defects, it is applicable to substitutional defect system since Raman features are less sensitive to the type of point defect.[20]

To establish a relationship between the slope of the equi-$n_D$ line on the $n_D$–$n_L$ plane ($A_D/A_{2D}$) and the measured defect parameters ($n_D$ or $L_D$), we plotted the slopes against $L_D^2$ (**Fig. 5b**). Given the linear relationship between $A_D/A_{2D}$ and $I_D/I_G$ (**Fig. 2c-d**), a similar linear relationship between the slope of the equi-$n_D$ line ($A_{2D}/A_D$) and $n_D^{-1}$ or $L_D^2$ is expected, consistent with the previous literature[17]. Based on this relationship (**Fig. 5b**), we derived the following equations connecting $A_{2D}$, $A_D$, $A_G$, and $L_D/n_D$:

$$\left(\frac{A_{2D}}{A_G}\right) = \frac{(L_D^2)^{1.13}}{92.1}\left(\frac{A_D}{A_G}\right) + 0.144, \quad L_D(nm) \quad (2)$$

$$\left(\frac{A_{2D}}{A_G}\right) = \frac{1}{92.1}\left(\frac{10^{14}}{\pi n_D}\right)^{1.13}\left(\frac{A_D}{A_G}\right) + 0.144, \quad n_D(cm^{-2}) \quad (3)$$

It is important to note that these equations are specific to the 532 nm (2.33 eV) laser excitation wavelength. Using these equations, $n_D$ can now be extracted from Raman spectra regardless of $n_L$. The universal form of the ratios $A_D/A_{D+2D}$ and $A_{2D}/A_{D+2D}$ was experimentally validated (**Extended Fig. 7**) using the experimental Raman data (**Extended Fig. 3**) and the derived equations. The crossing point of $A_D/A_{D+2D}$ and $A_{2D}/A_{D+2D}$ remained ~$7\times10^{12}$ cm$^{-2}$, independent of $n_L$. This relationship also holds for non-Bernal stacked bilayer graphene (**Extended Fig. 8**). Finally, we demonstrate the applicability of these equations by generating an $n_D$ map (**Fig. 5c**).

**Wavelength dependence of the Graphene Atlas.**

Because both $L_a$ and $L_D$ exhibit wavelength dependence in graphitic and graphenic materials[15,17], we investigated the influence of excitation laser energy ($E_L$) on the Raman data points within the $n_D$–$n_L$ plane. We focused on a specific graphene flake with two distinct $L_D$ values (5.05 and 8.20 nm; **Fig. 6a**) and performed Raman analysis using five different laser lines with $E_L$ ranging from 1.96 to 2.54 eV (**Fig. 6b**). Detailed Raman data for this flake are shown in **Extended Fig. 9**, and representative single-point Raman spectra are displayed in **Figs. 6b** and **6c**. As observed in other graphitic systems[15,17], we found laser energy-dependent variations in the intensity and peak position of D- and 2D-bands.

Since the two equi-$n_D$ lines corresponding to $L_D$ = 5.05 and 8.20 nm intersect in the third quadrant of the $n_D$–$n_L$ plane, we simplified the analysis by assuming that all equi-$n_D$ lines share a common intercept that varies with $E_L$. This assumption allows us to focus on the change of the slope of the equi-$n_D$ lines changing $E_L$. Using this constraint, we extracted the slopes $a(L_D, E_L^n)$ of the equi-$n_D$ lines for four laser excitations and two $L_D$ values using standardized major axis regression. To determine the appropriate exponent $n$ for $E_L$, we calculated the coefficient of variation (CV = $\sigma / \mu$, where $\sigma$ is the standard deviation and $\mu$ is the average) of $a/E_L^n$ for each $n$ (**Extended Fig. 10a**). While different CV minima were observed for the two different $L_D$ values, $n = 4$ yielded the minimum summed CV. This $E_L^4$ dependence is consistent with the observed wavelength dependence of $L_a$ and $L_D$ in previous studies.[15,17]

We also modeled the variation of the intercept with $E_L$ using an exponential relationship (**Extended Fig. 10b**). Combining $E_L$ dependence of the slope and the intercept, we derived the following general equation for arbitrary $L_D$ and $E_L$:

$$\left(\frac{A_{2D}}{A_G}\right) = \frac{(L_D^2)^{1.13} E_L^4}{2.71 \times 10^3}\left(\frac{A_D}{A_G}\right) + 2.43 \times 10^{-2}(E_L - 1.12)^{9.49} \quad (4)$$

This empirical equation accurately reproduces the Raman data obtained by specific laser excitation energies (dashed lines in **Fig. 6**).

**Conclusion**

We developed a universal and layer number-independent framework for quantifying point defects in graphenic materials, addressing a long-standing challenge in defect characterization. By preparing high-quality boron-doped graphenic materials with varying defect densities ($n_D$) and layer numbers ($n_L$) through mechanical exfoliation of boron-doped graphite, and subsequently analyzing them with Raman spectroscopy, we revealed a fundamental relationship between defect density and Raman spectral features. We demonstrated that the double-resonance-originated D-band and 2D-band represent competing processes, and their area ratio ($A_D/A_{2D}$) directly reflects $n_D$ and the inter-defect distance ($L_D$). Crucially, we found that graphenic materials with constant $n_D$ but varying $n_L$ exhibit a linear relationship on the $n_D$–$n_L$ plane, as the relative fractions of $A_D$ and $A_{2D}$ show a universal dependence on $n_D$, independent of $n_L$. This key finding led to the development of the *Graphene Atlas*, a powerful new tool for analyzing and quantifying defects in graphenic materials.

This work significantly advances the field of graphenic materials characterization by providing a universal, layer number-independent method for determining defect density. Unlike previous methods that are often limited to monolayer graphene or rely on intensity ratios affected by the number of layers, the *Graphene Atlas* uses area ratios, providing a more robust and accurate measure of $n_D$ in multilayer systems. This is especially important for applications where precise control over defect density is crucial, such as in electronic devices, sensors and catalysts. Because Raman spectroscopy is a versatile, powerful, and non-destructive technique, the *Graphene Atlas* enables real-time monitoring of defect evolution during a dynamic process such as growth, electrochemical reactions, and decomposition. This capability opens new avenues for understanding and optimizing these processes.

This framework sets the stage for application beyond graphene, including other 2D materials such as transition metal dichalcogenides and h-BN. Future work will integrate machine learning for automated defect mapping and predictive modeling, further enhancing its industrial applicability. These advancements will further solidify the *Graphene Atlas* as a valuable tool for accelerating nanomaterial innovation and development.

## Methods

### Boron Doping into Graphite

Kish graphite (Covalent materials) and boric acid ($H_3BO_3$, Wako Chemical) were used as starting materials and boron dopant precursor, respectively. Kish graphite flakes and 5wt% boric acid were placed into a graphite crucible; then the crucible was heated up under Ar flow using a graphite furnace (Kurata Giken) up to 2400°C. The temperature was held for 30 min followed by natural cooling to room temperature.

### Graphene Extraction

Mono-to-few layer graphene was extracted using a conventional mechanical exfoliation process. Both pristine and boron-doped graphite were subjected to the process, and scotch tape was used for mechanical exfoliation. After several times of exfoliation, the sticky side of the tape was attached to the $SiO_2$/Si substrate, followed by peeling off the tape, leaving graphene on the top of the substrate.

### Characterization

**Time-of-flight secondary ion mass spectroscopy**: Diffusion of the boron atoms into graphite is investigated by time-of-flight secondary ion mass spectroscopy (TOF.SIMS5, IONTOF). $Bi/Bi_3^{++}$ and $O_2$ were used as primary ion source and sputtering gun. A reflectron is set as positive mode to detect $^{10}B^+$ and $^{11}B^+$ ions.

**Raman spectroscopy**: Extracted mono-to-few layers of Gr or BGr were subjected to a Raman spectroscopy study. An inVia reflex Raman microscope (Renishaw) equipped with a 532 nm excitation laser, 1800 gr/mm grating, and x100 objective lens was mainly used to collect Raman spectra. Besides this, T64000 spectrometer (HORIBA Jobin Yvon) equipped with 488/515/532/561/633 nm laser excitation wavelengths, and 600/1800 gr/mm grating, and x100 objective lens was also used.

**Atomic force microscope:** The thickness and morphology of the pristine and boron-doped defective graphene were investigated using an atomic force microscope (E-Sweep, SII). Tapping mode was employed to avoid scratching the graphene.

**Data availability**

The data used in this study is available upon request from the corresponding author K.F.

**Acknowledgements**

K.F. acknowledges the financial support from the Kondo Memorial Foundation, and the Japan Society for the Promotion of Science (JSPS) KAKENHI Grant Number JP 22K04723. B.R.C. acknowledges the financial support from the Brazilian agencies CNPq, CAPES, and Brazilian Institute of Science and Technology in Carbon Nanomaterials (INCT-Nanocarbono). Y.A.K. acknowledges the financial support from the ITECH R&D program of the Ministry of Trade, Industry & Energy/Korea Evaluation Institute of Industrial Technology (MOTIE/KEIT) (RS-2023-00257573). M. T. would like to acknowledge the financial support from AFOSR (FA9550-23-1-0447).

**Author contribution**

**K.F.** designed and conceived this study, and conducted the experiments, measurements, data analysis, and prepared the manuscript. **B.R.C.** conducted Raman data analysis, and prepared the manuscript. **P.V.** performed the DFT calculations of Raman spectroscopy. **C.-S.K.** conducted AFM measurements. **Y.A.K.** contributed to the boron-doping experiment planning. **T.H.** and **M.T.** contributed to the experiment planning and the manuscript preparation. **K.F.**, **B.R.C.** and **M.T.** interpreted the results and wrote the manuscript with inputs from all authors. All authors discussed the results and contributed to the development of the manuscript.

**Competing interests**

The authors declare no competing interests.

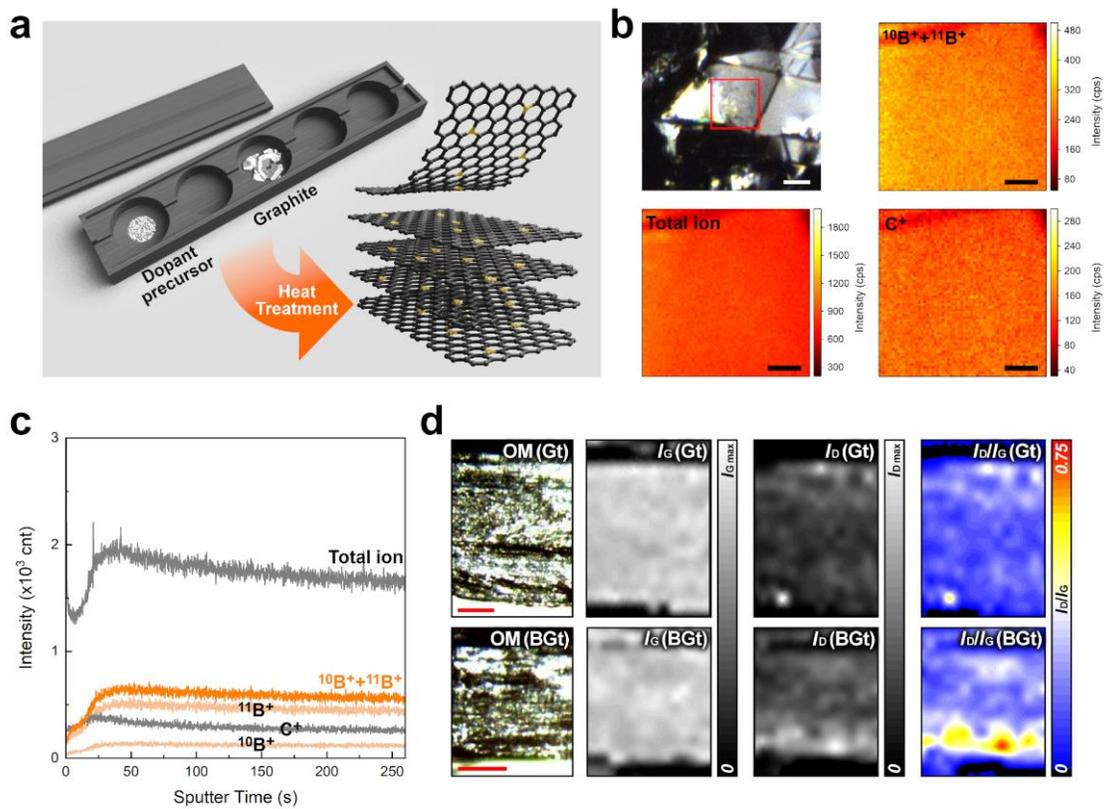

**Fig. 1 | Preparation and characterization of the boron-doped graphite. a**, Schematic illustration of the annealing setup and boron-doped graphite. A boron dopant precursor and graphite pieces are separately located. **b**, Optical image of the boron-doped graphite investigated by ToF-SIMS. Secondary ion images were collected from the area highlighted by the red rectangle. **c**, Depth profile of ion species obtained from the same area as (**b**). **d**, Cross-sectional optical images, and corresponding Raman mapping of pristine graphite (Gt) and boron-doped graphite (BGt).

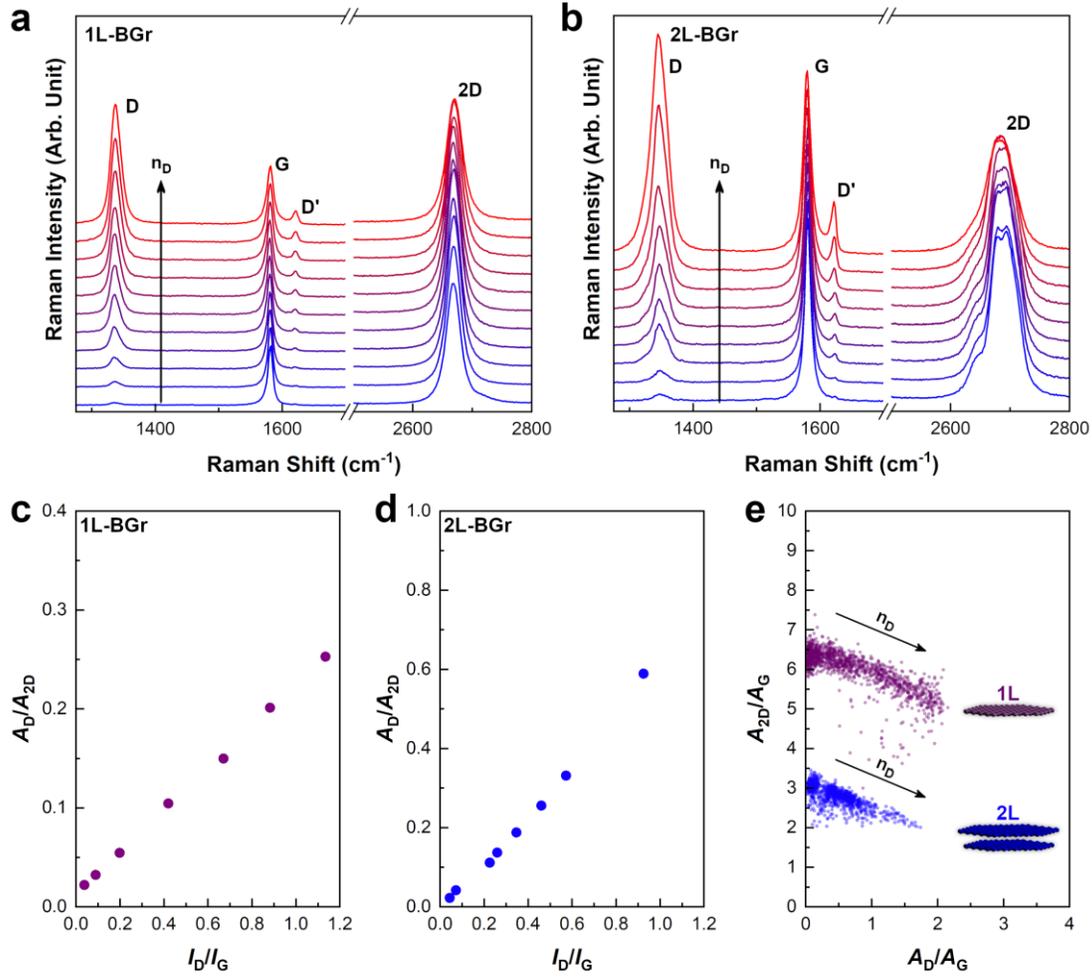

**Fig. 2 | Raman spectra of boron-doped monolayer and bilayer graphene. a-b**, Raman spectra of boron-doped monolayer graphene (1L-BGr), and bilayer graphene (2L-BGr), respectively. All spectra are stacked from low $n_D$ (blue) to high $n_D$ (red) based on relative D-band intensity. **c-d**, Relationship between $A_D/A_{2D}$ and $I_D/I_G$ in 1L- and 2L-BGr, respectively. **e**, Plotted Raman data points acquired from 1L-BGr (purple) and 2L-BGr (blue) plotted on the $A_D/A_G$ ($n_D$)–$A_{2D}/A_G$ ($n_L$) plane. Data points form a line with negative slope on $n_L$–$n_D$ plane.

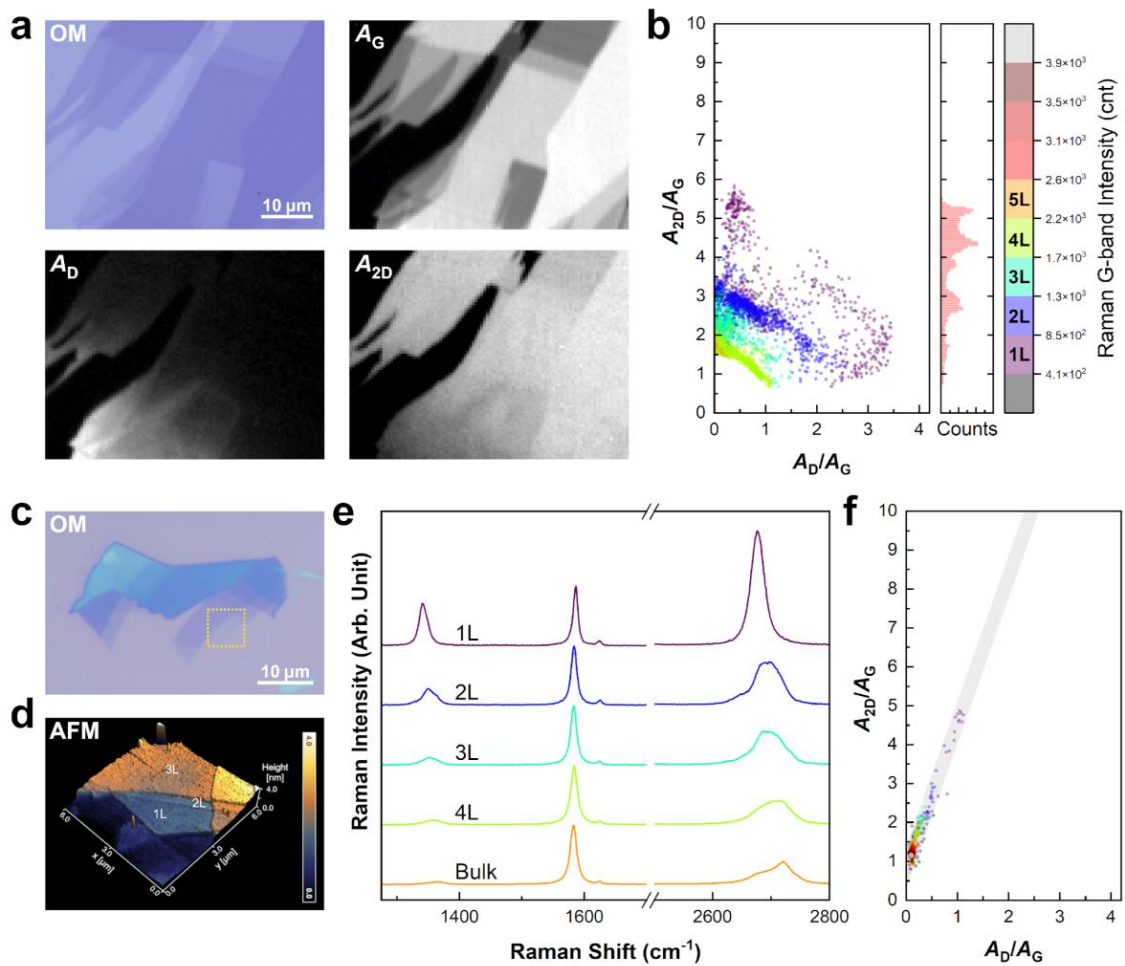

**Fig. 3 | Raman characterization of BGr single flake. a**, Optical microscopy (OM) image of BGr single flake and created Raman maps using $A_G$, $A_D$ and $A_{2D}$ (scale bar: 10 μm). **b**, Plotted Raman data acquired from BGr single flake shown in (**a**) on $n_D$–$n_L$ plane. All data points are colored using G-band intensity as shown in the histogram. **c-d**, Optical microscopy (OM) image (**c**), and atomic force microscopy image (**d**) acquired from rectangle highlighted in (**c**) of BGr single flake. **e**, $n_L$-dependent Raman spectra obtained from 1L–4L and bulky part in (**c**). **f**, Plotted Raman data acquired from BGr single flake shown in (**a**) on $n_D$–$n_L$ plane. All Raman data points align along a positively sloped line.

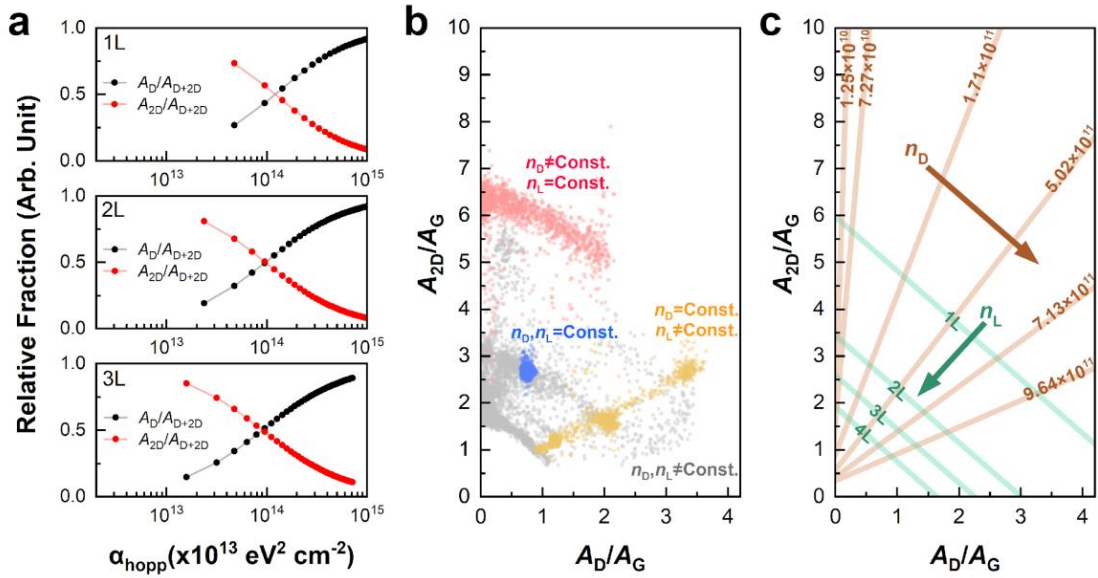

**Fig. 4 | Understanding the Raman data pattern on $n_D$–$n_L$ plane. a**, DFT-calculated Raman area $A_D$ and $A_{2D}$ for 1L-, 2L- and 3L-BGr and various $n_D$ and $α_{hopp}$. **b**, Categorization of Raman data pattern plotted onto $n_D$–$n_L$ plane. Four distinct Raman data patterns are represented in different colors. **c**, Guide for the $n_D$–$n_L$ plane. Equi-$n_D$ and equi-$n_L$ are colored in green and orange, respectively.

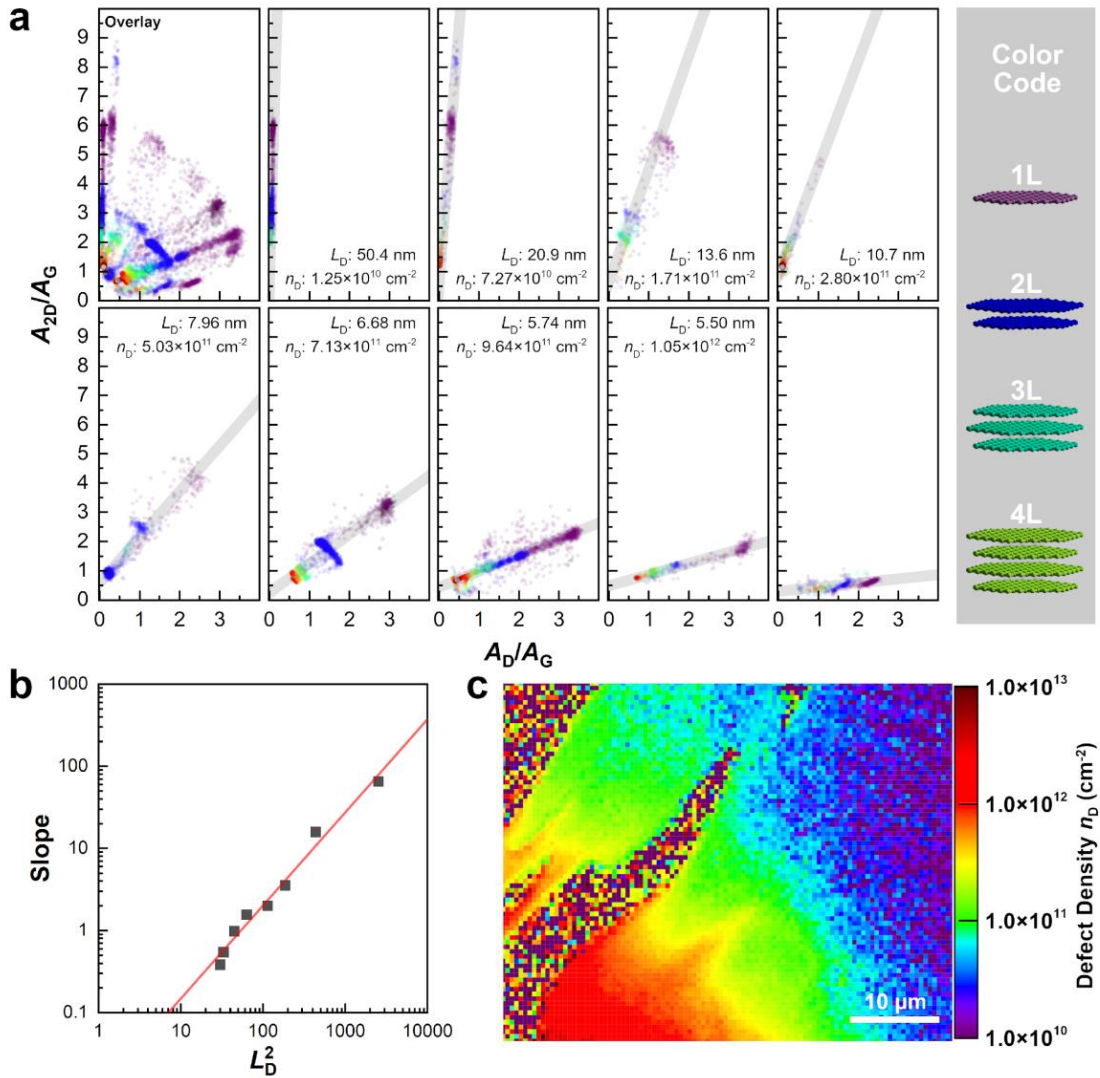

**Fig. 5 | Experimental calibration of equi-$n_D$ line in $n_D$–$n_L$ plane. a,** Raman $A_D/A_G$–$A_{2D}/A_G$ plot comparison of uniformly doped sample and non. $n_D$ and $L_D$ values for each equi-$n_D$ line were calculated using the monolayer region of the BGr single flake. **b**, Correlation of the slope for the equi-$n_D$ line and $L_D^2$. **c**, $n_D$ map obtained using the proposed equation, applied to **Figure 3a**.

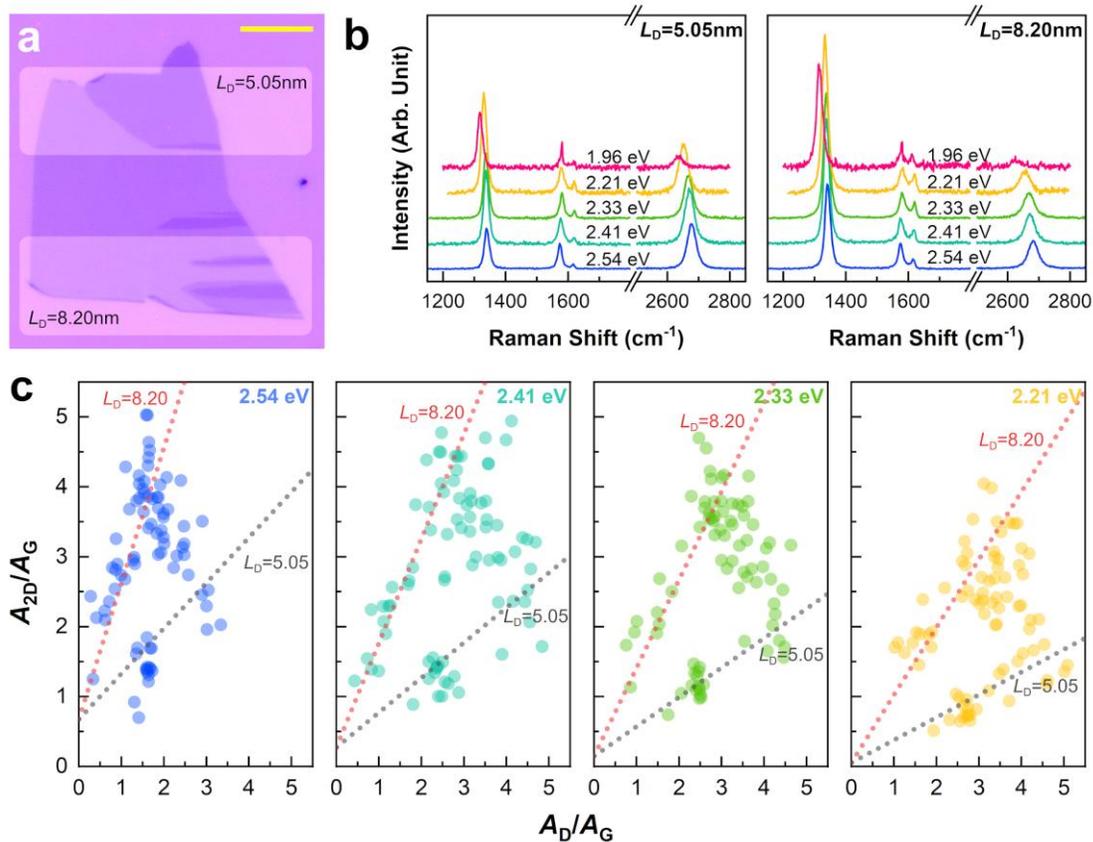

**Fig. 6 | Wavelength dependence of the $n_D$–$n_L$ plane. a**, Optical micrograph of graphene flake. Regions with two different $L_D$ are highlighted. The scale bar corresponds to 10 μm. **b**, Single point Raman spectra obtained using laser excitation energies of 1.96–2.54 eV. **c**, Raman data plotted onto the $n_D$–$n_L$ plane obtained with laser excitation energies of 2.21–2.54 eV. The linear lines are calculated results based on Eq. (4).

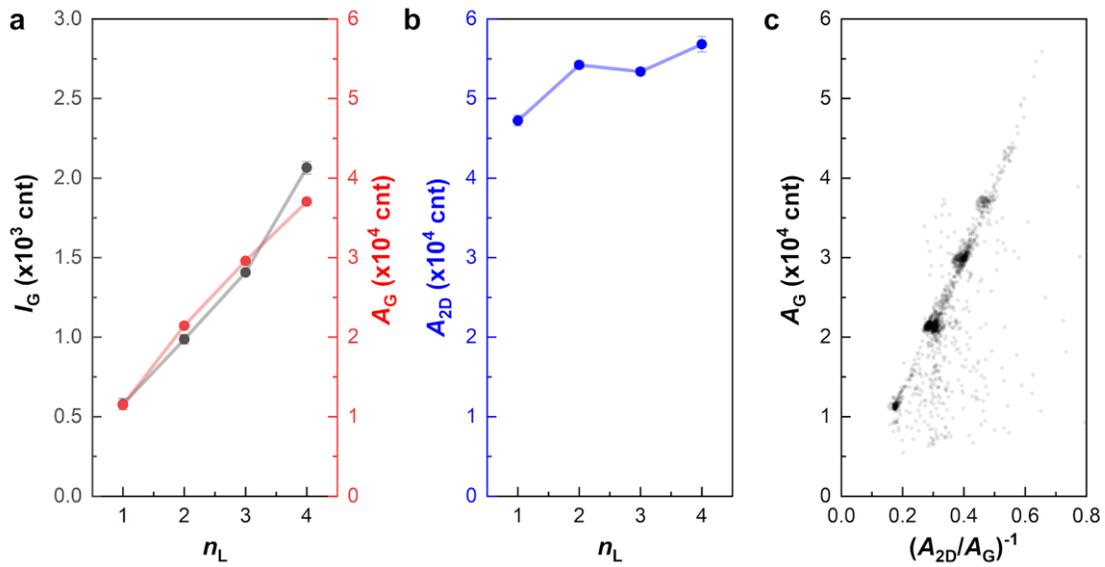

**Extended Data Fig. 1 | Layer number ($n_L$) dependency of the Raman features**. $I_G$, $A_G$ (**a**), and $A_{2D}$ (**b**) count plotted against the $n_L$. All data points are an average of 10 points extracted from the same graphene flake, and the error bar corresponds to the standard deviation. **c**, $A_G$ count plotted against $(A_{2D}/A_G)^{-1}$ correlates to $n_L$.

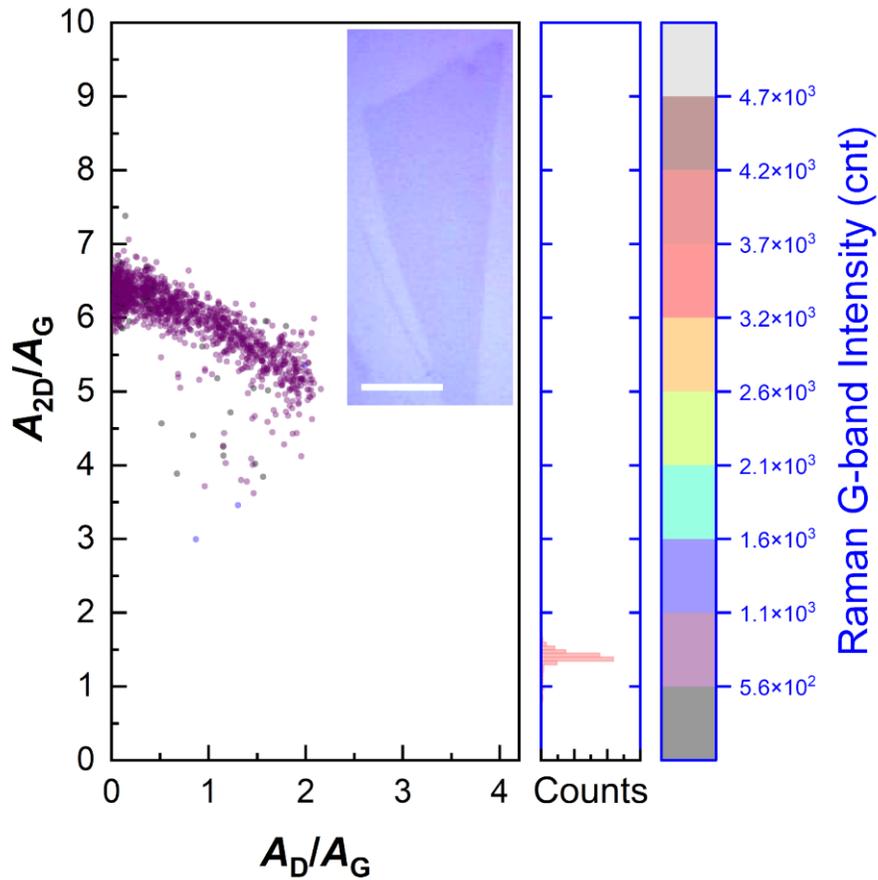

**Extended Data Fig. 2 |. Raman data with constant $n_L$ and non-constant $n_D$ plotted in the $n_D$–$n_L$ plane.** The inset shows an optical image of the graphene flake analyzed via Raman mapping (scale bar: 10 μm). Data markers in the plot are colored according to the G-band intensity ($I_D$). The uniform optical contrast in the optical image (inset) indicates that the examined graphene flake consists of only monolayer.

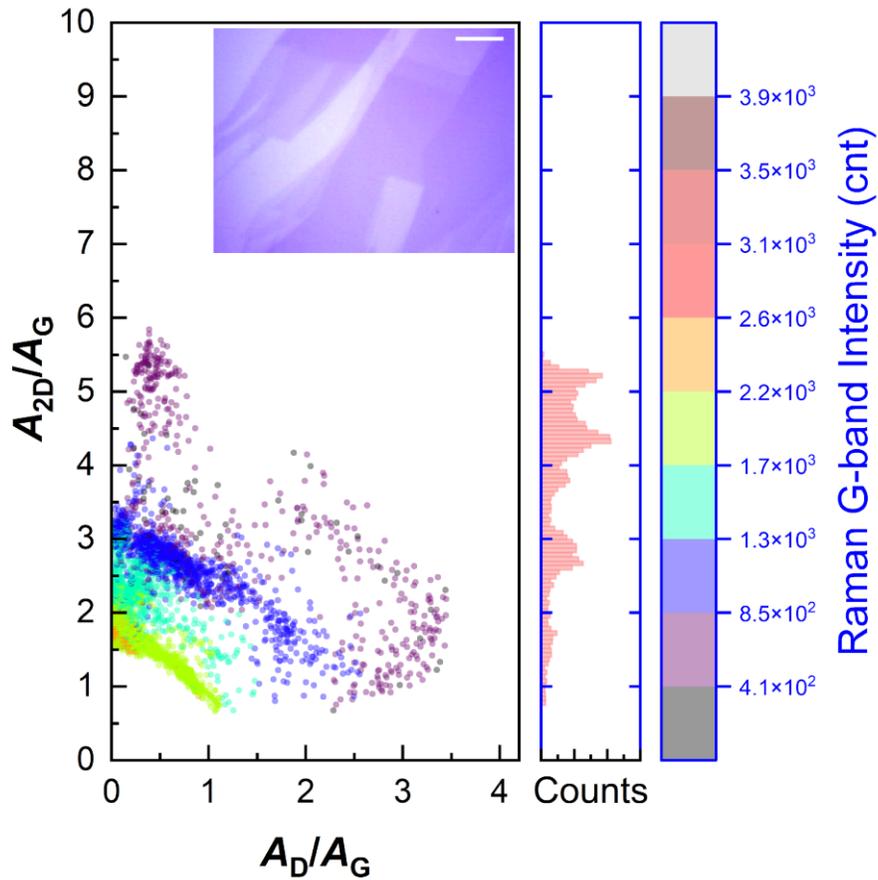

**Extended Data Fig. 3 | Raman data with non-constant $n_L$ and $n_D$ plotted in the $n_D$–$n_L$ plane**. The inset shows an optical image of the graphene flake analyzed via Raman mapping (scale bar: 10 μm). Data markers in the plot are colored according to the G-band intensity ($I_D$). The locally varying optical contrast in the optical image (inset) indicates that the examined graphene flake consists of 1-to-few layered graphene.

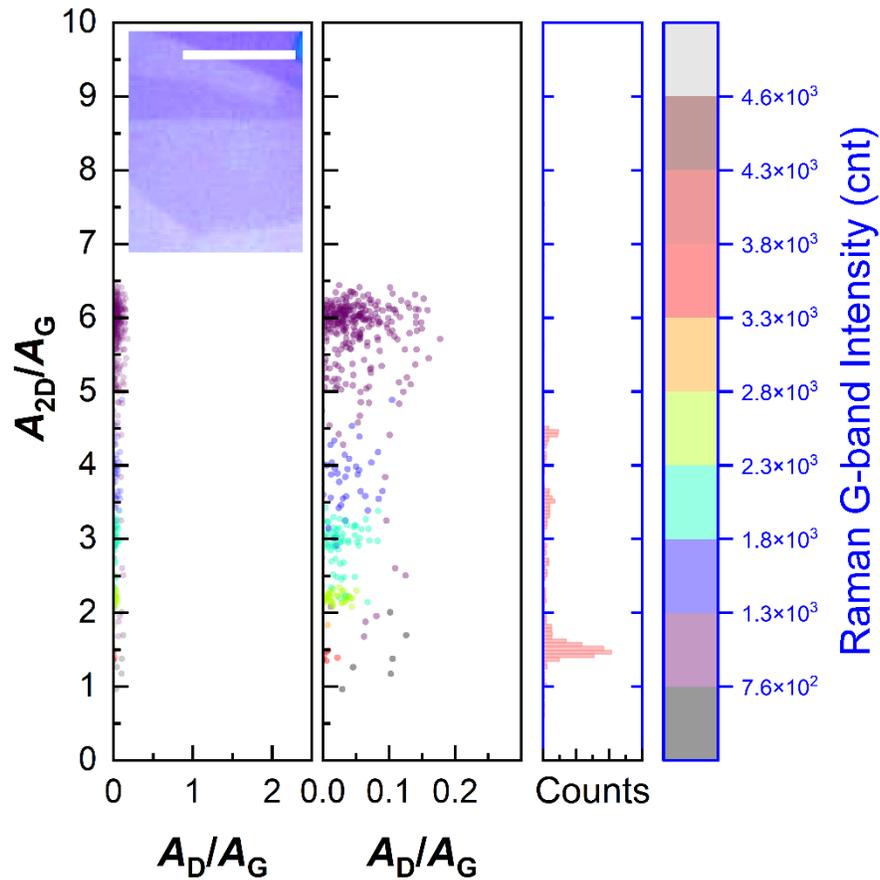

**Extended Data Fig. 4 | Raman data with non-constant $n_L$ and without boron dopants plotted in the $n_D$–$n_L$ plane.** The inset shows an optical image of the graphene flake analyzed via Raman mapping (scale bar: 10 μm). Data markers in the plot are colored according to the G-band intensity ($I_D$). The locally varying optical contrast in the optical image (inset) indicates that the examined graphene flake consists of 1-to-few layered graphene.

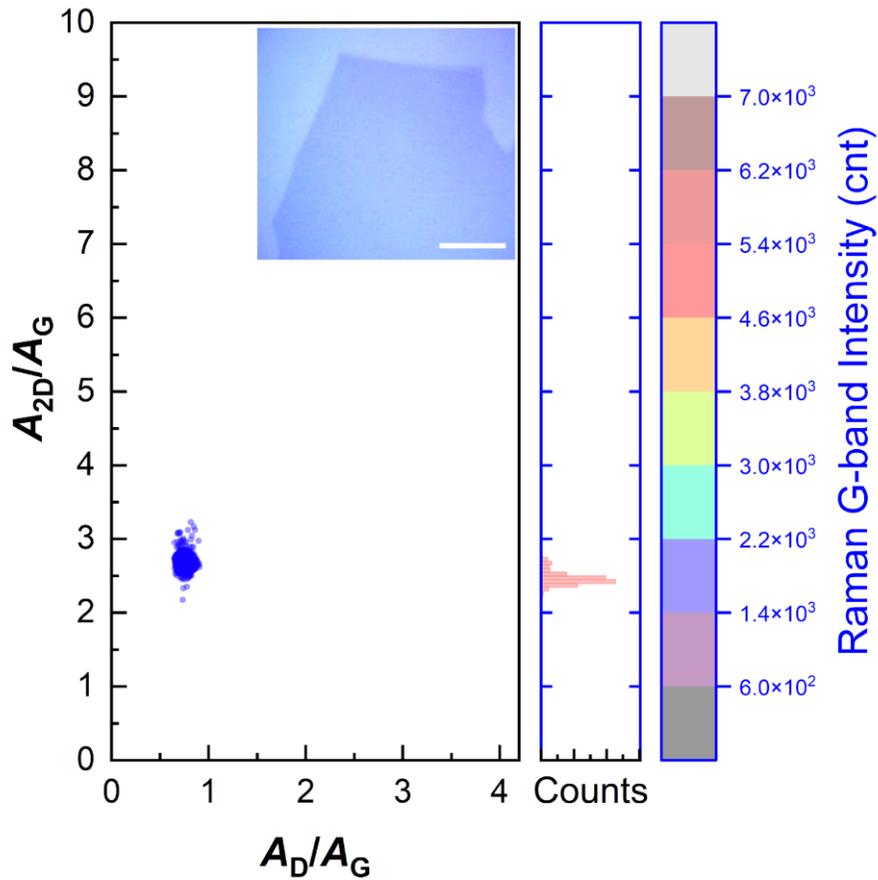

**Extended Data Fig. 5 | Raman data with constant $n_L$ and $n_D$ plotted in the $n_D$–$n_L$ plane.** The inset shows an optical image of the graphene flake analyzed via Raman mapping (scale bar: 10 μm). Data markers in the plot are colored according to the G-band intensity ($I_D$). The uniform optical contrast in the optical image (inset) indicates that the examined graphene flake consists of only bilayer graphene.

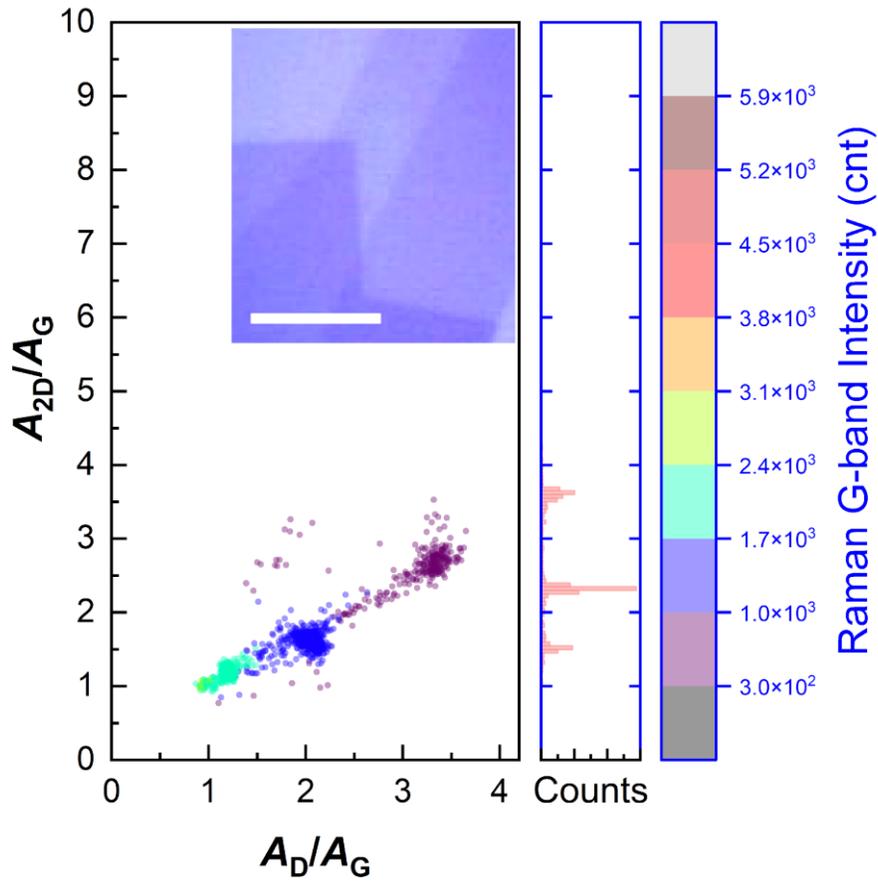

**Extended Data Fig. 6 | Raman data with non-constant $n_L$ and constant $n_D$ plotted in the $n_D$–$n_L$ plane**. The inset shows an optical image of the graphene flake analyzed via Raman mapping (scale bar: 10 μm). Data markers in the plot are colored according to the G-band intensity ($I_D$). The locally varying optical contrast in the optical image (inset) indicates that the examined graphene flake consists of 1-to-few layered graphene.

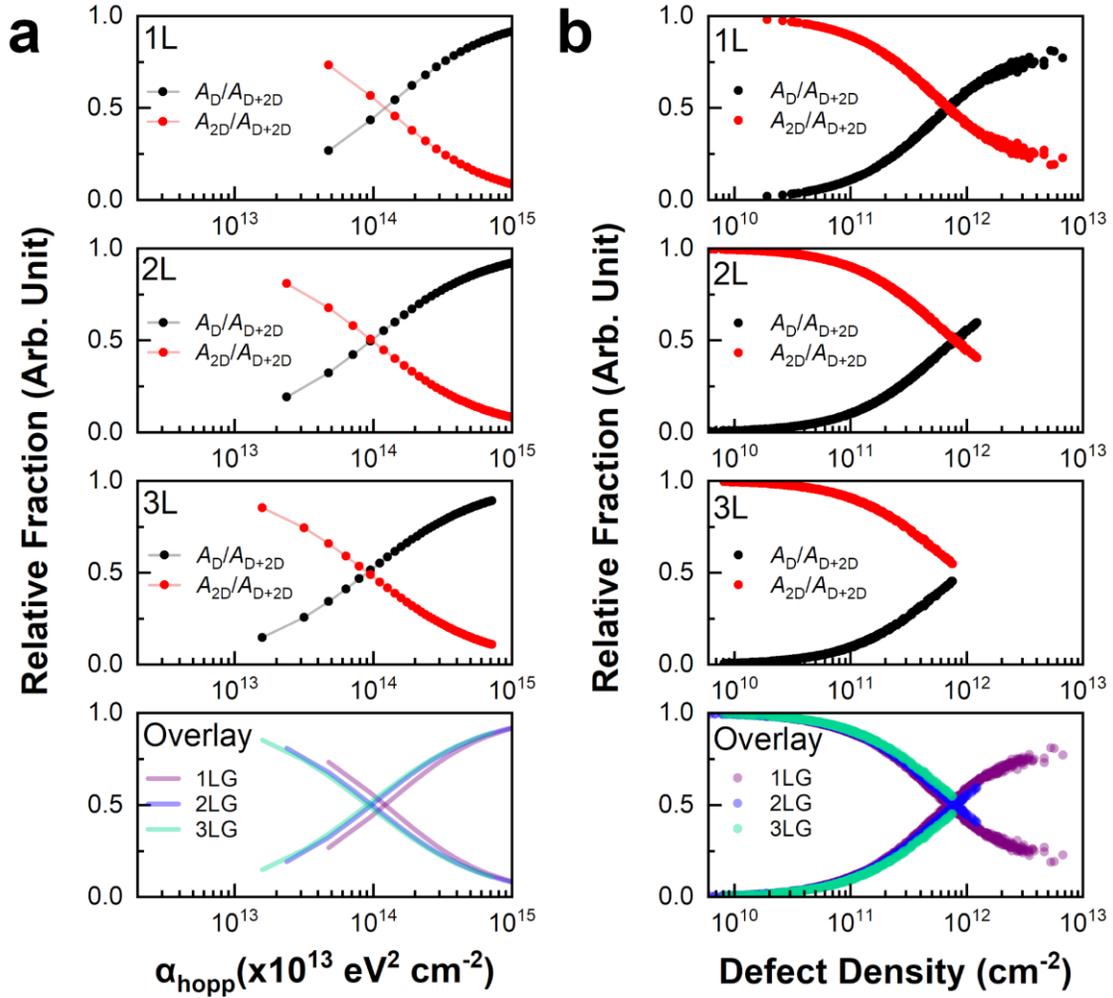

**Extended Data Fig. 7 | Universal form of the relative fraction of $A_D$ and $A_{2D}$. a**, DFT-calculated hopping parameter $\alpha_{hopp}$ dependency of the relative fraction of $A_D$ and $A_{2D}$ over $A_{D+2D}$. **b**, Experimentally reproduced $n_D$ dependency of the relative fraction of $A_D$ and $A_{2D}$ over $A_{D+2D}$ using Raman data shown in **Extended Fig. 3** and Eq. (3). Both theoretical and experimental results exhibited a universal form of the relative fraction of $A_D$ and $A_{2D}$ over $A_{D+2D}$ regardless of $n_L$.

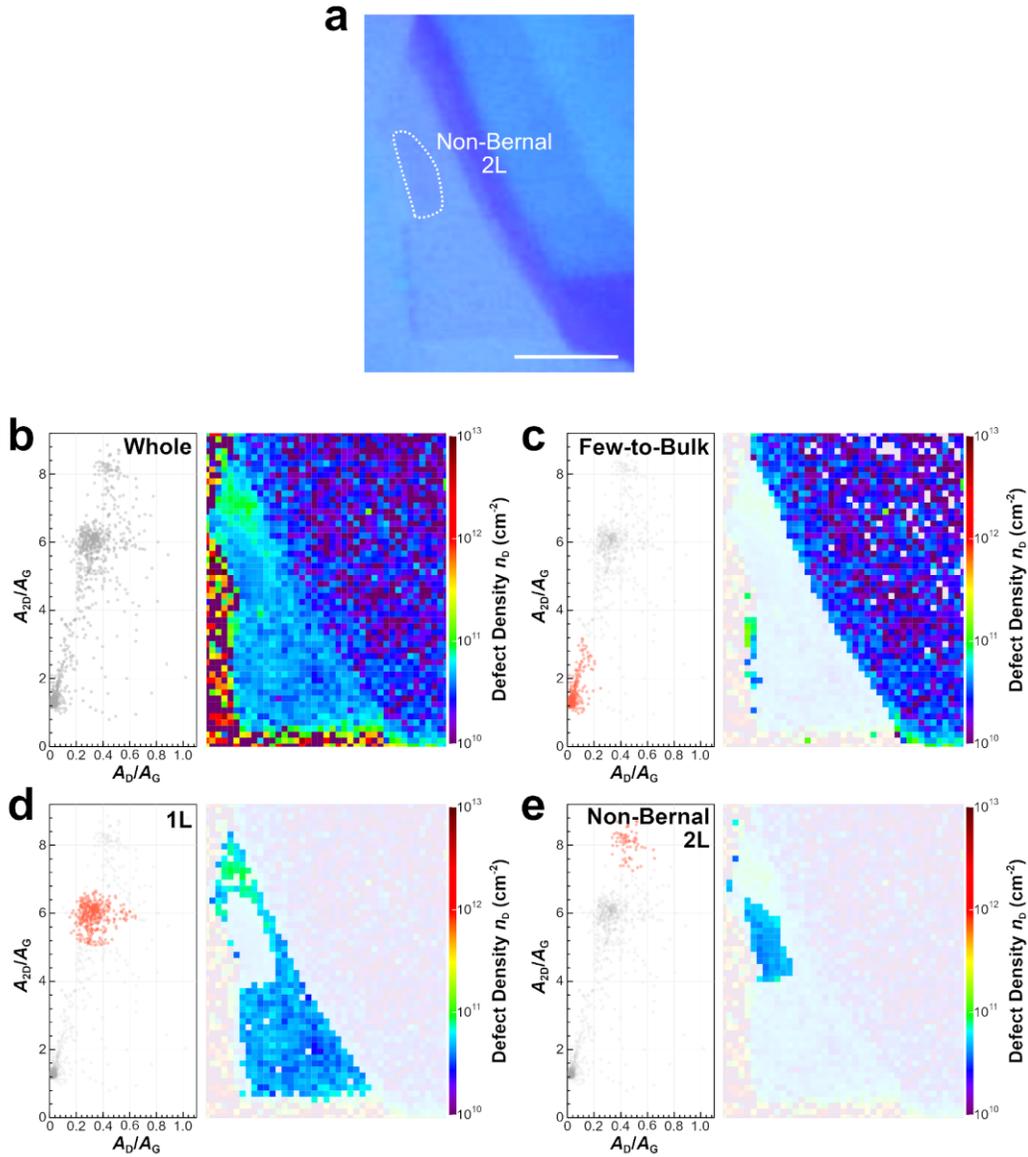

**Extended Data Fig. 8 | Graphene flake including non-Bernal 2L region. a**, Optical image of the graphene flake analyzed via Raman mapping (scale bar: 10 μm). **b-e**, Raman data plotted in $n_D$–$n_L$ plane (left-side frame) and Raman map (right-side frame) of whole graphene flake (**b**), Few-to-bulk region (**c**), 1L region (**d**), non-Bernal 2L region (**e**). The highlighted region of the graphene flake in the Raman map corresponds to the highlighted Raman data points in the $n_D$–$n_L$ plane.

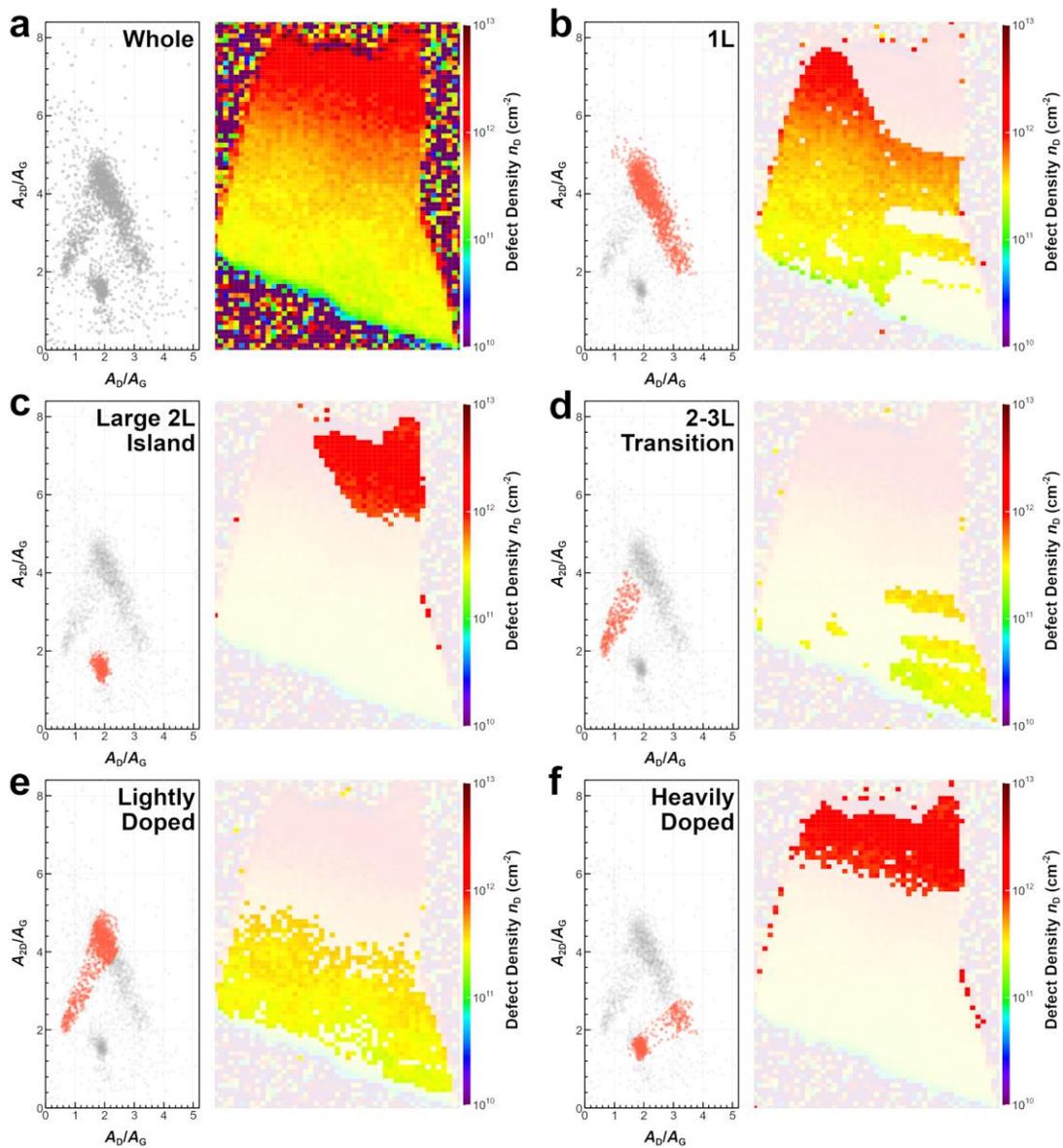

**Extended Data Fig. 9 | Graphene flake investigated by multi-wavelength Raman spectroscopy. a-f**, Raman data plotted in $n_D$–$n_L$ plane (left-side frame) and Raman map (right-side frame) of whole graphene flake (**a**), 1L-BGr region (**b**), large 2L-BGr island (**c**), 2L-3L transition region (**d**), lightly doped region (**e**), and heavily doped region (**f**). The highlighted region of the graphene flake in the Raman map corresponds to the highlighted Raman data points in the $n_D$–$n_L$ plane.

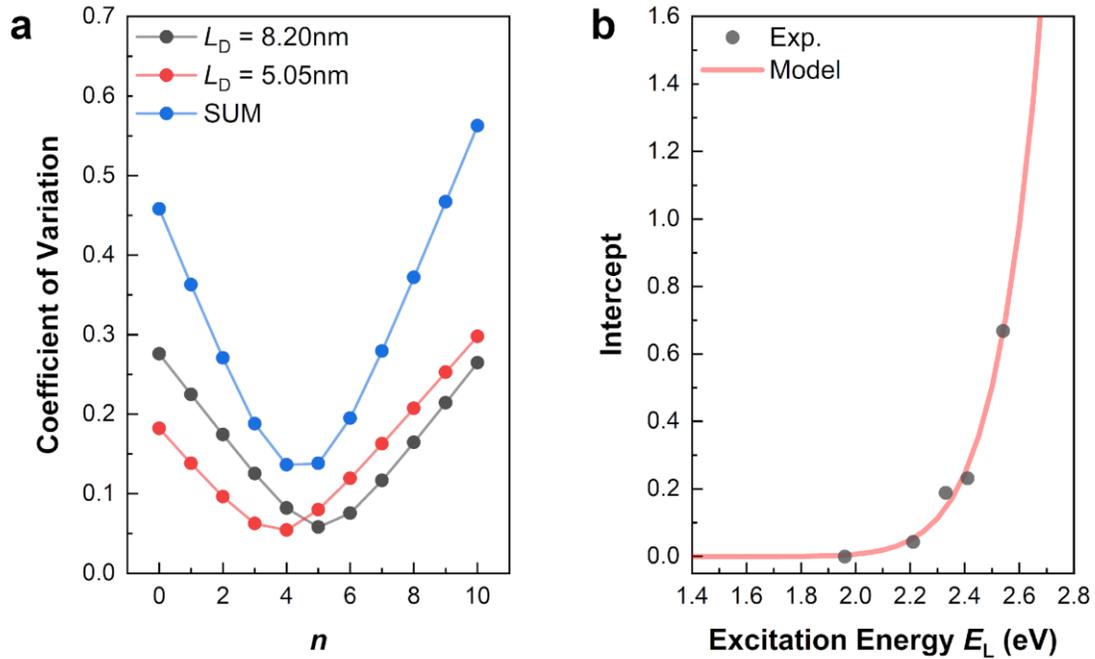

**Extended Fig. 10 | Fitting result to wavelength-dependent of equi-$n_D$ lines**. **a**, Coefficient of variation for slope *a* with $L_D$ equals to 5.05 nm and 8.20 nm. The summed coefficient of variation is also shown. **b**, $E_L$-dependent intercept variation, which is modeled with the exponential relationship. Note that two equi-$n_D$ lines are constrained to share a common intercept for simplification..

33